%% file: IGTMPchiribella.tex
\documentclass[a4paper]{jpconf}
\usepackage{amsmath,amssymb}
\def\Stset{\mathsf{St}}
\def\spc#1{\mathcal{#1}}
\def\map#1{\mathcal{#1}}
\def\>{\rangle}
\def\<{\langle}
\def\qed{$\blacksquare$}
\def\Lin{\mathsf{Lin}}
\def\Cmplx{\mathbb{C}}
\def\Irr{\mathsf{Irr}}
\def\UIrr{\mathsf{UIrr}}
\def\grp#1{\mathsf{#1}}
\def\Proof{{\bf  Proof.}~}
\def\d{{\rm d}}
\def\Supp{\mathsf{Supp}}
\newtheorem{lem}{Lemma}
\newtheorem{theo}{Theorem}
\newtheorem{cor}{Corollary}

\newtheorem{defi}{Definition}
\usepackage[matrix,frame,arrow]{xypic}\input{myQcircuit}
\begin{document}
\title{Group theoretic structures in the estimation of an unknown unitary transformation}
\author{Giulio Chiribella}
\address{Perimeter Institute for Theoretical Physics, 31 Caroline Street North, Waterloo, Ontario, N2L 2Y5, Canada}
\begin{abstract} This paper presents a series of general results about the optimal estimation of physical transformations in a given symmetry group.  In particular, it  is shown  how the different symmetries of the problem determine different properties of the optimal estimation strategy.  The paper also contains a discussion about the role of entanglement between the representation and multiplicity spaces  and about the optimality of square-root measurements.
\end{abstract}
\section{Introduction}

The estimation of an unknown unitary transformation in a given symmetry group is a general problem related to many stimulating topics, such as high-precision measurements \cite{metro}, coherent states and uncertainty relations \cite{aag}, quantum clocks \cite{clocks}, quantum gyroscopes \cite{refframe} and quantum reference frames \cite{review}. The aim of this paper is to provide a synthetic account of the general theory of optimal estimation for symmetry groups, presenting new proofs and highlighting the underlying group theoretical and algebraic structures into play.
\subsection{Prologue: dense coding}
Let us start with a simple example.  Suppose that we have at disposal a quantum system with two-dimensional Hilbert space  $\spc H = \Cmplx^2$ and a black box that performs an unknown transformation $\map U_{ij}, i,j \in\{0,1\}$, defined on the set $\Stset (\spc H)$  of all density matrices on $\spc H$ as $\map U_{ij} (\rho) :=  U_{ij} \rho U^\dag_{ij}, \forall \rho \in \Stset(\spc H)$, where $\{U_{ij}\}_{i,j=0}^1$ are the unitary matrices
\begin{equation*}
U_{00} =  \begin{pmatrix} 1 & 0 \\ 0&1 \end{pmatrix} , \quad  U_{01} =  \begin{pmatrix} 0 & 1 \\ 1&0 \end{pmatrix}, \quad U_{10} =  \begin{pmatrix} 1 & 0 \\ 0&-1  \end{pmatrix}  \quad U_{11} =  \begin{pmatrix} 0 & -1 \\ 1&0 \end{pmatrix}.
\end{equation*}
The  value of the indices $(i,j)$ is unknown to us and our  goal is to find it out using the black box only once.  Here the natural figure of merit is the minimization of the probability of error.

Clearly, if we apply the black box to a state $\rho \in\Stset(\spc H)$  there will always be an error:  the four states $\{\rho_{ij} :=  \map U_{ij} (\rho) \}_{i,j = 0}^1$ will never be perfectly distinguishable (in a two-dimensional Hilbert space there cannot be more than two perfectly distinguishable states!).
However, if we introduce a second system $\spc K \simeq \spc H$  and  apply the unknown transformation to the projector on the  entangled vector $|\Phi\>  =  \frac 1 {\sqrt 2} ( |0\>|0\> +   |1\>|1\>)  \in \spc H \otimes \spc K $, where $\{|0\>,|1\>\}$ is the standard basis for $\Cmplx^{2}$,  we obtain four states (the projectors on the vectors $|\Phi_{ij}\>  :=  (U_{ij} \otimes  I)  |\Phi\>$) that are perfectly distinguishable: $\< \Phi_{ij} | \Phi_{kl}\>    =  \Tr [U_{ij}^\dag U_{kl}] / 2 = \delta_{ik} \delta_{jl}$.
This remarkable fact was observed for the first time by Bennett and Wiesner, who exploited it to construct a protocol known as \emph{dense-coding} \cite{dense}.
The moral of dense coding is that  the use of entanglement with an ancillary system can improve dramatically the discrimination of unknown transformations.
Despite the extreme simplicity of the mathematics, this curious example points out structures that are much deeper than it might seem to the first sight.  The aim of the present paper is to illustrate these structures in the general context of the estimation of an unknown group transformation.
Our analysis will include dense coding, where the group of interest is  the  Klein group  $\mathbb Z_2 \times \mathbb Z_2$.

\section{General problem: estimation of an unknown group transformation}

\subsection{The problem}
Suppose that we have at disposal one use of a black box and that we want to identify the action of the black box.    Suppose also that we have some prior knowledge about the black box: in particular, we know that
\begin{enumerate}
\item the black box acts on a quantum system with finite dimensional Hilbert space $\spc H  \simeq  \Cmplx^d, d<\infty$.
\item it performs a deterministic transformation $\map U_g$ belonging to a given representation of a given symmetry group $\grp G$.
\end{enumerate}
Mathematically, a \emph{deterministic transformation}  (also known as \emph{quantum channel}) is described by a completely positive trace-preserving linear map $\map C$ acting on the set $\Stset(\spc H)$ of quantum states (non-negative matrices with unit trace) on the Hilbert space $\spc H$.  
The representation of the group $\grp G$ is then given a function $\map U: g  \mapsto \map U_g$  from $\grp G$ to the set of  quantum channels, with  the usual requirements
\begin{align*}
&\map U_e  = \map I_{\spc H} \qquad \\
&\map U_g  \map U_{g^{-1}}   = \map I_{\spc H} \qquad \forall g \in \grp G\\
&\map U_g  \map U_h  = \map U_{gh} \qquad \forall g,h\in\grp G
\end{align*}
where  $e \in \grp G$ is the identity element of the group and $\map I_{\spc H}$ is the identity channel, given by $\map I_{\spc H} (\rho)  := \rho, \forall \rho \in\Stset(\spc H)$.  In this paper the group $\grp G$ will be always assumed to be either finite or compact.

\subsection{Estimation strategies without ancillary systems}\label{subsec:simple}
Since the group $\grp G$ and the representation $\{\map U_g\}_{g \in \grp G}$ are both known, the problem here is to identify the group element $g\in \grp G$ that specifies the action of the black box.    How can we accomplish this task?  A first idea is to prepare the system in a suitable input state $\rho \in\Stset(\spc H)$ and to apply the unknown transformation $\map U_g$ to it, thus obtaining the output state  $\rho_g:=  \map U_g (\rho)$. The procedure can be represented diagrammatically as
\begin{equation*}
 \begin{aligned}
  \Qcircuit @C=1em @R=.7em @! R {& \prepareC {\rho} &\qw \poloFantasmaCn{\spc H} & \gate {\map U_g} & \qw \poloFantasmaCn {\spc H} &  \qw}
\end{aligned}  =   \begin{aligned} \Qcircuit @C=1em @R=.7em @! R {& \prepareC {\rho_g} &\qw \poloFantasmaCn {\spc H} &  \qw}\end{aligned}
\end{equation*}
In this way, the output state $\rho_g$ will carry some information about $g$ and we can try to extract this information with a quantum measurement and to produce an estimate $\hat g \in \grp G$.  The combination of the quantum measurement with our classical data processing can be described by a single mathematical object, namely a \emph{positive operator-valued measure}  (\emph{POVM}, for short).

Let us denote by $\Lin (\spc H)$ the set of linear operators on $\spc H$  and by $\Lin_+ (\spc H)\subset \Lin(\spc H)$ the set of non-negative operators.
If  $\grp G$ is a finite group, a POVM with outcomes in $\grp G$ is just a function $P:  \grp G \to  \Lin_+(\spc H)$  sending the element $\hat g \in \grp G$ to the non-negative operator $P_{\hat g} \in \Lin_+(\spc H)$ and satisfying the requirement $\sum_{\hat g \in \grp G}  P_{\hat g} = I_{\spc H}$, where $I_{\spc H}$ is the identity on $\spc H$.
The conditional probability of inferring $\hat g$ when the true value is $g$  is then given by the Born rule
$p(\hat g| g)  = \Tr[  P_{\hat g}  \rho_g]. $

In general, if $\grp G$ is a compact group and $\sigma (\grp G)$ is the collection of its measurable subsets \cite{sigmaalg}, a POVM with outcomes in $\grp G$ is an operator-valued measure $P :  \sigma (\grp G)  \to \Lin_+ (\spc H)$ sending the measurable subset $B \in\sigma (\grp G)$ to the non-negative operator $P_B \in \Lin_+ (\spc H)$ and satisfying the requirements
\begin{itemize}
\item  $P_\grp G = I_{\spc H}$
\item  if $B = \bigcup_{i =1}^\infty  B_i $ and  $\{B_i\}_{i =1}^{\infty} $ are disjoint, then $P_B  = \sum_{i=1}^\infty  P_{B_i}$.
\end{itemize}
The conditional probability that the estimate $\hat g$ lies in the subset $B \in\sigma(\grp G)$ is given by the Born rule
$p(B|g)  = \Tr[P_B  \rho_g].$

In the following I will frequently use the notation $P(\d \hat g)$ to indicate the POVM $P$.   Accordingly, I will also write $p(\d\hat g|g) =\Tr[P(\d \hat g)  \rho_g]$ and $P_B  =  \int_{B}   P(\d \hat g)$.  In the case of finite groups, the notation $P(\d \hat g)$  will be understood as synonym of $P_{\hat g}$ and the integral over a subset $B$ will be understood as a finite sum.
 The pair $(\rho, P)$ of an input state and a POVM will be referred to as an \emph{estimation strategy}.

\subsection{Estimation strategies with ancillary systems}
In the previous paragraph we discussed strategies where the unknown transformation was applied to a suitable state $\rho \in\Stset(\spc H)$.  However, these strategies are not most general ones:  we can introduce an ancillary system with Hilbert space $\spc K$,  prepare a \emph{bipartite} input state $\rho \in \Stset (\spc H \otimes \spc K)$, and then apply the black box on system $\spc H$, thus obtaining the output state $\rho_g: = (\map U_g  \otimes \map I_{\spc K}) (\sigma)$.  The schematic of this procedure is
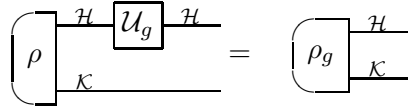
\begin{equation*}
 \begin{aligned}
  \Qcircuit @C=1em @R=.7em @! R
  {& \multiprepareC{1}{\rho} &\qw \poloFantasmaCn{\spc H} & \gate {\map U_g} & \qw \poloFantasmaCn {\spc H} &  \qw \\
  & \pureghost{\rho}    & \qw \poloFantasmaCn{\spc K}  &\qw &\qw &\qw  }
\end{aligned}  =   \begin{aligned} \Qcircuit @C=1em @R=.7em @! R {& \multiprepareC{1}{\rho_g} &\qw \poloFantasmaCn {\spc H} &  \qw\\
 & \pureghost{\rho_g}    & \qw \poloFantasmaCn{\spc K}  &\qw }\end{aligned}
\end{equation*}
In this case, to obtain an estimate $\hat g$ of the unknown group element $g$ we will use a POVM $P$ on the tensor product Hilbert space $\spc H \otimes \spc K$.

Classically, we would not expect any improvement from the use of an ancillary system because this system will not carry any information about the unknown transformation. However, here the state $\rho$ can be an entangled state, that is, a state that cannot be written as a convex combination of tensor product states.
As the example of dense coding teaches, the use of entanglement can improve the estimation dramatically.  Note, however, that mathematically every estimation strategy with ancillary system can be reduced to the form of an estimation strategy without ancillary system by choosing $\spc H' := \spc H \otimes \spc K$ and $\map U'_g  :=  \map U_g \otimes \map I_\spc K$.

\subsection{Cost of an estimation strategy}
To quantify how far is the guess $\hat g$ from the true value one typically introduces a \emph{cost function}  $c:  \grp G \times \grp G \to \mathbb R, (\hat g, g)  \mapsto c(\hat g, g)$ which takes its minimum value when $\hat g=g$.  If the true value is $g$, the expected cost of the estimation strategy $(\rho, P)$ is
\begin{align}
\nonumber c (\rho, P| g) &:= \int_{\grp G} ~ c(\hat g, g)  p(\d \hat g| g) =  \int_{\grp G} ~ c(\hat g, g)  \Tr[P(\d \hat g)  \rho_g].
\end{align}
Ideally, we would like to minimize the cost simultaneously for every $g$.  However, in general this is not possible because the number of states in the orbit $\{\rho_g\}_{g \in \grp G}$ is larger than the dimension of the Hilbert space and therefore there is no way in which the states $\{\rho_g\}_{g \in \grp G}$ can be perfectly distinguishable.

Since the true value $g \in \grp G$ is unknown, one approach to optimization is to minimize the \emph{worst-case cost} of the strategy, defined as
\begin{equation}
c_{wc}  (\rho, P)  =  \max_{g \in \grp G}  c(\rho, P| g).
\end{equation}
An alternative (and less conservative) approach consists in assuming a prior distribution $\pi (\d g )$ for the true value $g\in \grp G$ and in minimizing the average cost.  Given that $g$ is completely unknown, a natural choice of prior is the  Haar measure  $\d g$, normalized as $\int_{\grp G} \d g =1$.   With this choice, the average cost is given by
\begin{equation}
c_{ave}  (\rho, P)  :=  \int_{\grp G}  \d g ~  c(\rho, P|g).
 \end{equation}
For finite groups we can understand the integral as a finite sum and $\d g$ as synonymous of $1/|\grp G|$, where $|\grp G|$ is the cardinality of the group $\grp G$.

In general, the worst-case and the average approach are very different: for example, the functional $c_{ave} (\rho, P)$  is linear in both arguments while the functional $c_{wc}(\rho, P)$ is not.   The two approaches may lead to very different optimal strategies: for example, since $c_{ave}  (\rho, P)$ is linear in $\rho$ the optimal input state can be searched without loss of generality in the set of pure states $\rho  = |\varphi\>\<\varphi|$, while this reduction is generally not possible for the minimization of $c_{wc}  (\rho, P)$.    However, when the cost function enjoys  suitable group symmetries it is possible to show that there is a strategy  $(\rho,P)$ that is optimal for both the worst-case and the average approach. This point will be illustrated in Section \ref{sec:optest}.

\section{Basic notions and notations}
\subsection{Bipartite states}\label{subsec:bipartite}
Let $\{ |\phi_i \>  \}_{i=1}^{d_\spc H}$ (resp. $\{|\psi_j\>\}_{j=1}^{d_\spc K}$) be a fixed orthonormal bases for the Hilbert space $\spc H$ (resp.  $\spc K$).  The choice of orthonormal bases induces a bijective correspondence between linear operators from $\spc K$ to $\spc H$ and bipartite vectors in $\spc H \otimes \spc K$.   Following the ``double-ket notation" of Ref.  \cite{PLA}, if $A \in \Lin (\spc K, \spc H)$ is a linear operator from $\spc K$ to $\spc H$ we define the bipartite vector $|A\>\!\> \in\spc H \otimes \spc K$ as
\begin{equation}\label{kk}
|A\>\!\> : = \sum_{i=1}^{d_\spc H} \sum_{j = 1}^{d_\spc K}   \<  \phi_i |A  |\psi_j\>  ~ |\phi_i\>  |\psi_j\>.
\end{equation}
Two useful properties of this correspondence are given by the equations
\begin{align}
\<\!\<  A | B\>\!\>  &= \Tr [A^\dag B] \qquad  \qquad  \forall A, B \in \Lin (\spc K, \spc H)\\
\nonumber | A\>\!\>   &=  (A \otimes I_\spc K)  |I_\spc K\>\!\>  \\
 \phantom{| A\>\!\> }&=(I_\spc H \otimes A^T)  |I_\spc H \>\!\> \qquad \forall A \in \Lin (\spc K,\spc H), \label{modconj}
\end{align}
where $A^T$ denotes the transpose of $A$ with respect to the fixed bases.
Using the singular value decomposition of $A$, it is easy to see that the state $|A\>\!\>$  is of the product form $|A\>\!\>  = |\phi\>|\psi\> $ if and only if $A$ is rank-one.  Moreover, if the dimension of $\spc K$ is larger than $d_{\spc H}$,  then for every  state $|A\>\!\>$ there exists a $d_{\spc H}$-dimensional subspace  $\spc K_A$ such that $|A\>\!\> \in   \spc H \otimes \spc K_A$.     This is clear from Eq. (\ref{modconj}): the dimension of the image of $A^T$ cannot exceed the dimension of its domain $\spc H$.

\subsection{Group representations}\label{subsec:grouprep}
The unknown transformation in our estimation problem belongs to a group representation $\{\map U_g\}_{g \in \grp G}$, where each map $\map U_g$ is a completely positive trace preserving map sending operators in $\Lin (\spc H)$  to operators in $\Lin (\spc H)$.  Now, the condition $ \map U_{g^{-1}} \map U_g  = \map I_\spc H$ implies that the map $\map U_g$ must have the form $\map U_g (\rho)  = U_g \rho U_g^\dag $, where $U_g\in\Lin (\spc H)$ is a unitary matrix.  In particular, for $g=e$ we can choose $U_e = I_\spc H$.  Moreover, the condition $\map U_g \map U_h  = \map U_{gh} , \forall g,h \in \grp G$ implies that the unitaries $\{  U_g\}_{g \in \grp G}$ define a projective unitary representation, that is, a function $U:  g \mapsto U_g $   sending elements of the group to unitary operators and satisfying the relation  $U_g U_h  = \omega (g,h)  U_{gh}, \forall g,h \in \grp G$, where $\omega (g,h) \in \Cmplx$ is a multiplier, satisfying the properties
\begin{equation*}
\begin{array}{lr} |\omega (g,h)|=1  \qquad  &\forall g,h \in \grp G\phantom{.}\\
\omega (g,h)  \omega (gh,k)= \omega (g,hk)  \omega (h,k) \qquad & \forall g,h,k \in \grp G\phantom{.}\\
 \omega (g,e) = \omega (e,g)  = 1 \qquad &\forall g \in \grp G.
 \end{array}
\end{equation*}

Since the group $\grp G$ is compact and the space $\spc H$ is finite dimensional, with a suitable choice of basis the Hilbert space can be decomposed as
\begin{equation}\label{cgh}
\spc H = \bigoplus_{\mu \in\Irr (U)}     \spc H_\mu  \otimes \Cmplx^{m_\mu} ,
\end{equation}
where the sum runs over the set $\Irr (U)$ of all irreducible representations contained in the isotypic decomposition of $U$, $\spc H_\mu$ is a \emph{representation space}  of dimension $d_\mu$, carrying the irreducible representation $U^\mu$, and $\Cmplx^{m_\mu}$ is a \emph{multiplicity space}, where $m_\mu$ is the multiplicity of the irreducible representation $U^\mu$ in the decomposition of $U$.
Accordingly, the representation $U$ can be written in the block diagonal form
\begin{equation}\label{cg}
U =  \bigoplus_{\mu \in \Irr (U)}  U^{\mu}  \otimes I_{m_\mu} .
\end{equation}
 Note that all irreducible representations $U^\mu$ must be projective unitary representations and must have the same multiplier $\omega$.

In the optimization of the estimation strategy $(\rho, P)$ we can take the multiplicities $m_{\mu}$  as large as we want.  Indeed, introducing an ancillary system $\spc K$
 means replacing the Hilbert space $\spc H$ with the tensor product  $\spc H' : =\spc H \otimes \spc K$ and replacing the unitary $U_g$  with the unitary  $U_g'  := U_g \otimes I_{\spc K}$ for every $g \in \grp G$.    In Eqs. (\ref{cgh}) and (\ref{cg})  this means replacing the multiplicity space $\Cmplx^{m_\mu}$ with $\Cmplx^{m_\mu} \otimes \spc K$ and the multiplicity $m_\mu $ with $m_\mu' := m_\mu d_{\spc K}$.   In particular, we are free to choose the ancillary system $\spc K$ so that the condition $m_\mu\ge d_\mu$ is met for every $\mu \in \Irr (U)$.

\subsection{Bipartite states and group representations}
Here we combine the facts observed in the paragraphs \ref{subsec:bipartite} and \ref{subsec:grouprep}.
Let us consider a vector  $|\Psi\>\in \spc H$ in a Hilbert space where the representation $U$ acts.  Using the isotypic decomposition of Eq. (\ref{cgh}) and the correspondence of Eq. (\ref{kk}) for each tensor product $\spc H_\mu \otimes \Cmplx^{m_\mu}$, we can write
\begin{equation*}
|\Psi\>  =  \bigoplus_{\mu \in \Irr(U)}  |\Psi_\mu\>\!\> \in \bigoplus_{\mu \in \Irr (U)} \spc H_\mu \otimes \Cmplx^{m_\mu},
\end{equation*}
where $\Psi_\mu $ is a linear operator in $\Lin(\Cmplx^{m_\mu}, \spc H_\mu)$.  In other words, the vector $|\Psi\>$ can be written as a linear superposition of bipartite vectors. As we will see in the next sections, the fact that we can have entanglement between each representation space $\spc H_{\mu}$ and the corresponding multiplicity space $\Cmplx^{m_\mu}$  is the key ingredient to the construction of the optimal estimation strategies.


Suppose that  $U$ has been chosen so that $m_\mu \ge d_\mu$ for every $\mu \in \Irr (U)$.    From the last observation  of paragraph \ref{subsec:bipartite} we know that each vector $|\Psi_\mu\>\!\>$ will be contained in a subspace $\spc H_\mu  \otimes \spc K_\mu$, where $\spc K_\mu\simeq \Cmplx^{d_\mu} \simeq \spc H_\mu$ is a suitable subspace of $\Cmplx^{m_\mu}$.  Therefore, every vector $|\Psi\>$  belongs to a suitable subspace $\widetilde{\spc H}    \simeq  \bigoplus_{\mu \in \Irr (U)}  \spc H_\mu \otimes \spc H_\mu $.  Note that the whole orbit $\{|\Psi_g\>  :=  U_g  |\Psi\>\}_{g \in \grp G}$ belongs to the subspace $\widetilde {\spc H}$.
In conclusion, as long as we are interested in \emph{pure} input states we can effectively replace $\spc H$ with the space $\widetilde{\spc H}  := \bigoplus_{\mu \in \Irr (U)} \spc H_\mu \otimes \spc H_\mu$  where $m_\mu  = d_\mu$ for every $\mu \in  \Irr (U)$ and we can replace the representation $U$ with the representation $\widetilde U$ defined by $\widetilde U_g :=  \bigoplus_{\mu \in \Irr (U)}  U_g^\mu  \otimes I_{\mu}$, where $I_\mu$ denotes the identity on $\spc H_\mu$.

\subsection{The class states}
An important family of states in $\widetilde {\spc H} = \bigoplus_{\mu\in\Irr (U)}  \spc H_\mu \otimes \spc H_\mu$ is the family of states of the form
\begin{equation}\label{classstates}
|\Phi\>  =  \bigoplus_{\mu \in \Irr( U)}  \frac{c_\mu}{\sqrt{d_\mu}}  |I_\mu \>\!\>,  \qquad c_\mu \in \Cmplx, \sum_{\mu\in\Irr (U)}  |c_\mu|^2 =1
\end{equation}
These states are linear superpositions of maximally entangled states in the sectors $\spc H_\mu \otimes \spc H_\mu$.  In the following I will call these states \emph{class states}.  The reason for the name comes from the fact that these states correspond to \emph{class functions} (see definition below)  through the Fourier-Plancherel theory (see e.g. \cite{folland}).  The remaining part of this paragraph is aimed at making this correspondence clear.

Let us start from Fourier-Plancherel theory. Denote by $\Irr (\grp G, \omega)$ the set of all  irreducible representations of $\grp G$ with multiplier $\omega$ and denote by $u^\mu_{ij} (g) :  =  \<\mu,i  |   U^\mu_g  |\mu,j\>$  the matrix element of the operator  $U^\mu_g$ with respect to a fixed orthonormal basis $\{|\mu,i\>\}_{i =1}^{d_\mu}$ for the representation space $\spc H_\mu$.  Recall that the functions  $\{  \tilde u_{ij}^{\mu^*} (g):= \sqrt{d_\mu} u_{ij}^{\mu*}  (g)  \}_{\mu \in \Irr (\grp G, \omega); i,j =1, \dots, d_\mu}$ are an orthonormal basis for the Hilbert space $L_2 (\grp G, \d g)$ endowed with the scalar product $\<f_1, f_2\>  := \int_{\grp G} \d g ~ f_1^* (g)  f_2(g)$.     Using this fact, we can expand every function $f \in L_2 (\grp G, \d g)$ as
\begin{equation}\label{expansion}
|f\>  = \bigoplus_{\mu \in \Irr (\grp G,\omega)} \sum_{i,j =1}^{d_\mu}   f^\mu_{ij}  |\tilde u_{ij}^{\mu*}\>  .
\end{equation}
On the Hilbert space $L_2(\grp G, \d g)$ we consider the action of $\grp G$ given by the left-regular representation  $T$ with multiplier $\omega$, defined by
$(T_g  f) (h)  :=  \omega (g,g^{-1} h)  f (g^{-1}  h), \quad \forall f \in L_2(\grp G, \d g) ,  \forall g,h \in \grp G.$ \cite{dirac}.
In particular, the transformation of the functions $\tilde u_{ij}^{\mu*}(g)$ is given by
\begin{equation}\label{sonno1}
T_g |\tilde u_{ij}^{\mu*} \>  = \sum_{k=1}^{d_\mu}  u_{ki}^{\mu}  (g)  |\tilde u_{kj}^{\mu*}\>,
\end{equation}
that is, for every fixed value of $j$ the vectors $\{|\tilde u_{ij}^{\mu*}\>\}_{i=1}^{d_\mu}$ span an invariant subspace carrying the irreducible representation $U^\mu$.
It is then easy to see that the unitary operator  $ V :  L_2(\grp G, \d g)  \to  \bigoplus_{\mu \in \Irr (\grp G,\omega)} \spc H_\mu \otimes \spc H_\mu$ defined by
\begin{equation}\label{sonno2}
 V |u_{ij}^{\mu^*}\> : =  |\mu,i\>|\mu,j\> \qquad \forall \mu \in \Irr (\grp G, \omega), \forall i,j=1, \dots, d_\mu
 \end{equation}
 intertwines the two representations $T_g$  and $W_g  = \bigoplus_{\mu \in \Irr (\grp G, \omega)}  U_g^\mu \otimes I_\mu$: indeed, using Eqs. (\ref{sonno1}) and (\ref{sonno2})  one has $ V  T_g  |u_{ij}^{\mu*} \>  =  U_g^{\mu} |\mu,i\>  |\mu,j\>= (U^\mu_g \otimes I_\mu) V  |u_{ij}^{\mu*} \>$ for every $\mu\in\Irr (\grp G, \omega),\forall  i,j=1, \dots, d_\mu$, and therefore  $VT_g = W_g V$.
Applying the unitary $V$ to the expansion of Eq. (\ref{expansion}) we then obtain
\begin{equation*}
V  |f\>  = \bigoplus_{\mu \in \Irr (\grp G, \omega)}  \sum_{i,j=1}^{d_\mu}  f_{ij}^{\mu}  |\mu,i\>|\mu,j\>.
\end{equation*}

Let us come now to class functions.  A class function is a function that is constant on conjugacy classes, that is, $f (hgh^{-1}) = f(g)$ for every $g,h\in \grp G$.   Any class function can be written as a linear combination of irreducible characters, namely $f (g)  = \sum_{\mu \in \Irr (\grp G)}  f_\mu  \chi_{\mu}^*(g)$, where $\chi_{\mu}^*(g)  =  \Tr [ U_g^{\mu*}]$.
Exploiting the correspondence given by the unitary $V$ we obtain
\begin{equation*}
V |f\>  =  \bigoplus_{\mu \in \Irr (\grp G,\omega)}   \frac{f_\mu}{\sqrt {d_\mu}}  | I_\mu\>\!\>.
\end{equation*}
It is now clear that the class states of Eq. (\ref{classstates}) are nothing but the projection of the class functions to the finite dimensional subspace $\widetilde {\spc H}  \subset \bigoplus_{\mu \in \Irr (\grp G,\omega)}  \spc H_\mu \otimes \spc H_\mu$.

 Besides providing a justification for the choice of the name \emph{class states},  the Fourier-Plancherel theory also provides a hint of the fact that the class states are optimal for estimation.
Indeed, if we take the scalar product of two states in the orbit $\{|\Phi_g\>:=  \widetilde U_g |\Phi\>\}$ we obtain
$\< \Phi_g  |  \Phi_h\>  = \sum_{\mu \in \Irr (U)}  \frac {|c_\mu|^2}{d_\mu}   \chi_{\mu} (g^{-1} h).  $
 If $c_\mu$  is chosen so that $c_\mu  =  \lambda d_\mu$ for some constant $\lambda \in \Cmplx$ we then obtain $\<\Phi_g|  \Phi_h\>  \propto \sum_{\mu  \in\Irr(U)}  d_\mu  \chi_\mu (g^{-1} h) $, which is nothing but the truncated version of the Dirac delta $\delta  (g, h)  = \sum_{\mu \in \Irr (\grp G, \omega)} d_\mu  \chi_\mu (g^{-1}  h)$.
 This provides an heuristic argument for the optimality of the class states, although the choice of the coefficients $\{c_\mu\}$ to provide the ``best"  finite dimensional approximation of the Dirac delta will depend on the choice of the cost function $c(\hat g, g)$.
\subsection{Notations}
If $  V \in \Lin (\spc H)$ is a unitary matrix, the symbol $\map V$ will always denote the completely positive map $\map V (\rho) = V \rho V^\dag$.
Let $\map C: \Lin (\spc H)  \to \Lin (\spc K)$ be a completely positive map and let $\map C(\rho)  =  \sum_{i=1}^r  C_i \rho C_i^\dag$, $C_i \in \Lin (\spc H, \spc K)$  be a Kraus form for $\map C$.
The notation $\map C^\dag, \map C^*, \map C^T$ will be used for the following maps
\begin{align*}
\map C^\dag (\rho) &: = \sum_{i=1}^r   C_i^\dag \rho C_i\\
\map C^* (\rho) & := \sum_{i=1}^r   C_i^* \rho C_i^T\\
\map C^T (\rho) &: = \sum_{i=1}^r  C_i^T \rho C_i^* =  \left(\map C^*\right)^\dag (\rho),
\end{align*}
where the complex conjugation $*$ and the transpose $T$ are defined with respect to the fixed bases for $\spc H$ and $\spc K$. Note that the map $\map C^\dag$ is the adjoint of $\map C$ with respect to the Hilbert-Schmidt scalar product $\<\!\<A|B\>\!\> = \Tr[A^\dag B], A, B, \in\Lin (\spc H)$: indeed, we have $\<\!\<  A| \map C(B\>\!\> =  \<\!\<  \map C^\dag (A)| B\>\!\>.)$  Note also that the definition of the maps $\map C^\dag, \map C^*, \map C^T$ does not depend on the choice of a particular Kraus form.


Consider the Hilbert space $\widetilde {\spc H}  =\bigoplus_{\mu \in \Irr (U)}  \spc H_{\mu} \otimes \spc H_\mu$, the representation $  \widetilde U  =  \bigoplus_{\mu \in \Irr (U)}   U^\mu \otimes I_\mu$, and let
\begin{align*}
\widetilde U' & = \left\{  \bigoplus_{ \mu \in \Irr (U)}   I_\mu \otimes B_\mu ~|~  B_\mu\in\Lin (\spc H_\mu) \right\} \\
\widetilde U'' &= \left\{  \bigoplus_{\mu \in \Irr (U)}    A_\mu   \otimes I_\mu~|~  A_\mu \in \Lin (\spc H_\mu) \right\}
\end{align*}
be the commutant and the bicommutant of $\widetilde U$, respectively. Every class state in Eq. (\ref{classstates}) then defines a modular conjugation: for every operator $A =  \bigoplus_{\mu \in \Irr (U)}    A_\mu   \otimes I_\mu\in \widetilde U''$ I will denote by $A^R\in \widetilde U'$  its modular conjugate, given by
\begin{equation}
A^R := \bigoplus_{\mu \in \Irr(U)}  I_\mu \otimes A_\mu^*.
\end{equation}
Every class state $|\Phi\> \in \widetilde {\spc H}$ then enjoys the symmetry
\begin{equation}\label{symmetryclassstates}
\widetilde U_g \widetilde U_g^R  |\Phi\>  =  |\Phi\> \qquad \forall g \in \grp G,
\end{equation}
as it can be easily verified using Eq. (\ref{modconj}) for every $\mu \in \Irr(U)$.   The  modular conjugate representation will be denoted by $\widetilde U^R$.    For a completely positive map $\map C (\rho) = \sum_{i=1}^r  C_i \rho  C_i^\dag$ with $C_i  \in \widetilde U''$ for every $i=1, \dots, r$ we will denote by $\map C^R$ the completely positive map  $\map C^R (\rho):= \sum_{i=1}^r  C_i^R  \rho C_i^{R\dag}$.   Note that in general we have
\begin{equation}\label{sguscia}
\map C (|\Phi\>\<\Phi|)  = \map C^{R\dag}  (|\Phi\>\<\Phi|).
\end{equation}
Introducing the \emph{swap map} $\map S$, defined by $\map S \left( \bigoplus_{\mu \in \Irr (U)}  A_\mu \otimes B_\mu    \right) :=  \bigoplus_{\mu \in \Irr (U)}  B_\mu \otimes A_\mu $, we obtain the relations
\begin{align}\label{swap}
A  &=  \map S\left(  A^{R*}\right)  \qquad \forall A \in \widetilde U''\\
\map C & =  \map S  \map C^{R*}  \map S \qquad \forall  \map C :  \map C(\rho) = \sum_{i =1}^r  C_i \rho C_i^\dag, \quad C_i \in \widetilde U''~ \forall i =1,\dots,r.
\end{align}

\section{Optimal estimation}\label{sec:optest}
\subsection{Left-invariant cost functions: optimality of covariant measurements}
In this paragraph I will shortly review two classic results about the structure of covariant measurements and about their optimality. The proofs of these results can be found in the monographs \cite{helstrom,holevo,belavkin}.

Let us start from the definition:
\begin{defi}
A POVM $ P: \sigma (\grp G)  \to \Lin_+(\spc H)$ is \emph{covariant} with respect to the representation  $\{U_g\}_{g \in \grp G}$ if
$P_{g B}  = \map U_g  (P_B) \quad \forall g \in \grp G, \forall B \in\sigma (\grp G)$.
 \end{defi}

Covariant POVMs have a very simple structure:
\begin{theo}[Structure of covariant POVMs]
Let $\grp G$ be a compact group, $U$ be a projective unitary representation of $\grp G$ on the Hilbert space $\spc H$.
A POVM  $P: \sigma (\grp G)  \to \Lin_+ (\spc H)$ is covariant with respect to $U$  if and only if it has the form
\begin{equation}\label{covpovm}
P(\d \hat g) =   \map U_{\hat g}  (\xi)   ~  \d \hat g
\end{equation}
where $\d \hat g$ is the normalized Haar measure and $\xi\in\Lin_+(\spc H)$ is a suitable operator, called the \emph{seed} of the covariant POVM.
\end{theo}

The special form of covariant POVMs provides a great simplification in optimization problems, because it reduces the optimization of the whole POVM to the optimization of a single non-negative operator $\xi\in\Lin_+ (\spc H)$.   Luckily, this simplification can be done without loss of generality in most situations. Precisely, covariant measurement are optimal whenever the cost function $c(\hat g, g)$ is \emph{left-invariant}, that is $c(h \hat g, h g) = c(\hat g, g)  \forall g,h \in \grp G$.

\begin{theo}[Optimality of covariant POVMs]\label{theo:optcovpovm}
Let $\grp G$ be a compact group, $\{U_g\}_{g \in \grp G}$ be a unitary projective representation of $\grp G$ on the Hilbert space $\spc H$ and $c(\hat g,g)$ be a left-invariant cost function. Then, for every input state $\rho\in\Stset (\spc H)$ the optimal  POVM $P$ for both the worst-case and uniform average approach can be assumed without loss of generality to be covariant.  With this choice one has $c_{ave} (\rho,P)  =c_{wc} (\rho, P) = c(\rho,P| g ) \forall g \in \grp G $.
\end{theo}

\subsection{Right-invariant cost functions: optimality of  class states}
In this paragraph I illustrate with a new proof a recent result on the optimality of class states \cite{entest}. This result is dual to the classic result on the optimality of covariant POVMs showed in the previous paragraph.

A central feature of quantum theory is the validity of the \emph{purification principle} \cite{purification}:  for every mixed state $\rho \in \mathsf{St} (\spc H)$ there exists a Hilbert space $\spc K$ and a pure state $|\Psi\>\!\>  \in \spc H \otimes \spc K$ such that  $\rho = \Tr_{\spc K}[ |\Psi\>\!\>\<\!\<\Psi|]$
The pure state $\Psi$ is referred to as a \emph{purification} of $\rho$.  From Eq. (\ref{modconj})  we have $\rho  =  \Psi\Psi^\dag$ and $\Supp (\rho) = \mathsf{Rng} (\Psi)$, where $\Supp$ and $\mathsf{Rng}$ denote the support and the range, respectively. The simplest example of purification of a state $\rho\in\Stset(\spc H)$ is the \emph{square-root purification}  $|\rho^{\frac 12}  \>\!\> \in\spc H\otimes \spc H'$, where $\spc H' \simeq \spc H$.

A large number of quantum features, such as teleportation and no cloning,  are simple consequences of the purification principle \cite{purification}. The structure of the optimal states for the estimation of group parameters is no exception to that.  In the following I will give a simple proof of the optimality of the states in Eq. (\ref{classstates}) based on the purification principle. To this purpose I will use a remarkable consequence of purification, namely the fact, first noticed by Schr\"odinger \cite{schro}, that every ensemble decomposition of a state can be induced from the purification via a quantum measurement on the purifying system:
\begin{lem}\label{lem:steer}
Let $(X, \sigma (X))$  be a measurable space and let $\rho (\d x)  :  \sigma  (X)  \to  \Lin_+ (\spc H),    B   \mapsto  \rho_B$ be an ensemble, that is, an operator-valued measure such that $\rho_X \in \Stset (\spc H)$.    If    $|\Psi\>\!\> \in \spc H \otimes \spc K$ is a purification of $\rho_X$  then there exists a POVM $P  :  \sigma (X)  \to \Lin_+ (\spc K) , B \in\sigma(X) \mapsto P_B $  such that
\begin{equation}
\rho_B = \Tr_{\spc K} [   (I_{\spc H} \otimes P_B) |\Psi\>\!\>\<\!\< \Psi |   ] \qquad \forall B \in \sigma (X).
\end{equation}
\end{lem}
{\bf Proof.}  Let $\Psi^{-1}$ be the inverse of $\Psi$ on its support and define the POVM $P$ via $P_B: = \left[\Psi^{-1}   \rho_B  (\Psi^\dag)^{-1}\right]^T$. The POVM $P$ provides a resolution of the projector on the range of $\Psi^T$ and can be easily extended to a resolution of the identity on $\spc K$.  Using the identity $\Psi \Psi^{-1}= P_{\Supp (\rho_X)}$ where $\Pi_{\Supp(\rho_X)}$ is the projector on the support of $\rho_X$  we have
\begin{align*}
 \Tr_{\spc K} [  (I_{\spc H} \otimes P_B) |\Psi\>\!\>\<\!\< \Psi |    ] &=   \Tr_{\spc K} [   (I_{\spc H} \otimes   (\Psi^{-1} \rho_B)^T    |\Psi\>\!\>\<\!\< \Psi | (I_{\spc H} \otimes (\Psi^*)^{-1}  )  ]\\
  &= \Tr_{\spc H'} [   (  |\Psi\Psi^{-1}  \rho_B\>\!\>\<\!\< \Psi\Psi^{-1} |   ]\\
& = \Tr_{\spc H'}[ |\Pi_{\Supp(\rho_X)}  \rho_B \>\!\> \<\!\<  \Pi_{\Supp (\rho_X) } ]\\& = \rho_B.
\end{align*}
\qed

Suppose that  the cost function $c(\hat g,g)$ is \emph{right-invariant}, that is $c(\hat g h, gh) = c(\hat g,g) , \forall \hat g,g ,h\in \grp G$. Then we have the following
\begin{theo}[Optimality of the purification of invariant states]\label{theo:purinv}
Let $\grp G$ be a compact group, $U$ be a  projective unitary representation of $\grp G$ on the Hilbert space $\spc H$ and $c(\hat g,g)$ be a right-invariant cost function. Then, the optimal input state for both the worst-case and uniform average approach can be assumed without loss of generality to be the purification of an invariant state $\rho \in U'$.  Denoting by  $|\Psi\>\in\spc H \otimes \spc L$ such a purification, there exists an optimal POVM $P$ on $\spc H \otimes \spc L$ such that  $c_{ave} (\Psi, P)  = c_{wc} (\Psi,P) = c(\Psi, P|g), \forall g \in \grp G$.
\end{theo}
\Proof  Suppose that  $\sigma\in \Stset(\spc H\otimes \spc K)$ is an optimal input state and that $Q (\d \hat g ) $ is an optimal POVM for the average approach (resp. for the worst case approach). We now prove that there is a Hilbert space $\spc L$ and another state $|\Psi\> \in \spc H\otimes \spc L$ that is the purification of an invariant state $\rho\in U'$ and is optimal as well.
Consider the orbit $\{\sigma_h  : = (\map U_h \otimes \map I_{\spc K}) (\sigma)\}_{h \in \grp G}$ and the operator-valued measure $\sigma (\d h)  = \sigma_h  \d h$.   Clearly the state
$\sigma_{\grp G}  = \int_{\grp G}  \d  h~   \sigma_h$
 is invariant under $U \otimes I_\spc K$.  Let $ |\Psi\> \in \spc H\otimes \spc K \otimes \spc H' \otimes \spc K'$, with $\spc H'\simeq \spc H, \spc K' \simeq \spc K$,  be the square-root  purification of $\sigma_{\grp G}$ and define $\spc L := \spc K \otimes \spc H'\otimes \spc K'$.   Note that $|\Psi\> \in \spc H \otimes \spc L$  is also a purification of the invariant state $\rho : = \Tr_{\spc K}  [ \sigma_\grp G]\in U'$. We now have to show that $|\Psi\>$ is optimal for the uniform average (resp.  worst-case) approach.    By lemma \ref{lem:steer}, we know that there is a POVM $ M_h \d h$ on $\spc H' \otimes \spc K'$ such that
$ \rho_h  = \Tr_{\spc H' \otimes\spc K'} [   (I_{\spc H} \otimes I_\spc K  \otimes M_h) |\Psi \> \< \Psi |   ] \quad \forall h \in\grp G. $
Define now the POVM  $P (\d \hat g)  :=  \int_{\grp G}  \d h ~ Q(\d (\hat g h))   \otimes  M_h $.   It is easy to see that the average cost of the strategy $(\Psi, P)$ is equal to the average cost of the strategy $(\sigma, Q)$:
\begin{align*}
c_{ave}  (\Psi, P )  & =\int_{\grp G}  \d g  \int_{\grp G}    ~ c(g , \hat g)   ~  \<\!\<  \Psi_g|   P(\d \hat g)    |\Psi_g\>\!\>\\
& =\int_{\grp G}  \d g    \int_{\grp G}  \d h   \int_{\grp G}  ~  c(g , \hat g)   ~  \<\!\<  \Psi|  (  U_g^\dag Q(\d (\hat gh))  U_g   \otimes M_h   ) |\Psi\>\!\>\\
&   =\int_{\grp G}  \d h \int_{\grp G} \d g    \int_{\grp G}   ~ c(g , \hat g)   ~   \Tr[ U_g^\dag  Q(\d (\hat gh))   U_g      \sigma_h] \\
&   =\int_{\grp G}  \d h  \int_{\grp G}  \d ( g h) \int_{\grp G} ~  c(gh , \hat gh)   ~  \Tr[     Q(\d (\hat gh))   \sigma_{gh}]\\
&   =c_{ave} (\sigma,Q),
\end{align*}
having used the right-invariance of the Haar measure  and of the cost function. On the other hand, the worst-case cost of the strategy $(\Psi, P)$ cannot be larger than the worst-case cost of the strategy $(\sigma,Q)$:
\begin{align*}
c_{wc}  (\Psi, P )  & =\max_{g \in \grp G }  \int_{\grp G}    ~ c(\hat g ,  g)   ~  \<\  \Psi_g|   P(\d \hat g)    |\Psi_g\>\\
& =\max_{g \in \grp G}    \int_{\grp G}  \d h   \int_{\grp G}  ~  c( \hat g,g)   ~  \<\ \Psi|  (  U_g^\dag Q(\d (\hat gh))  U_g   \otimes M_h   ) |\Psi'\>\\
&   = \max_{g \in \grp G} \int_{\grp G}  \d h    \int_{\grp G}   ~ c(g , \hat g)   ~ \Tr [   U_g^\dag  Q(\d (\hat gh))   U_g   \sigma_h]\\
&   =\max_{g \in \grp G} \int_{\grp G}  \d h  \int_{\grp G} ~  c( \hat gh,gh)   ~  \Tr    [Q(\d (\hat gh))   \sigma_{gh}]\\
&   =\max_{g \in \grp G} \int_h \d h ~ c(\sigma,Q|gh) \\
&   \le  \int_h \d h  \max_{g \in \grp G}~ c(\sigma,Q|gh) \\
& = c_{wc} (\sigma,Q).
\end{align*}
Hence, we proved that, both for the average and for the worst case approach, whichever the optimal strategy $(\sigma, Q)$ is, the strategy $(\Psi, P)$ cannot be worse.  Finally, we observe that for the strategy  $(\Psi,P)$ we have
 $c (\Psi, P |g)     =\int_{\grp G}  \d h  \int_{\grp G} ~  c( \hat gh,gh)   ~  \Tr[    Q(\d (\hat gh))    \sigma_{gh} ] = \int_h \d h ~ c(\sigma,Q|h)$.
This proves that $c(\Psi,P|g)$ is independent of $g\in \grp G$, whence $c(\Psi,P|g) = c_{ave} (\Psi,P) = c_{wc}(\Psi,P)$.
\qed

\begin{theo}[Optimality of the class states]
Let $\grp G$ be a compact group, $U$ be a unitary projective representation of $\grp G$ on the Hilbert space $\spc H$ satisfying the property  $d_\mu \le m_\mu \forall \mu \in \Irr(U)$ and $c(\hat g,g)$ be a right-invariant cost function. Then, the optimal input state for both the worst-case and uniform average approach can be assumed without loss of generality to be a  class state of the form
\begin{equation}\label{optinp}
|\Phi\> =\bigoplus_{\mu \in \Irr  (U) }   \sqrt{\frac{p_\mu}{d_\mu}}   |  I_{\mu}\>\!\> \in \widetilde{\spc H}  = \bigoplus_{\mu \in \Irr(U)} \spc H_\mu \otimes \spc H_\mu \subseteq \spc H,
\end{equation}
where $p_\mu\ge 0$ and $\sum_{\mu} p_\mu=1$.
\end{theo}
\Proof By theorem \ref{theo:purinv}, we know that there exists an optimal state that is the purification of an invariant state $\rho \in U'$.  Now the form of an invariant state is
$\rho  = \bigoplus_{\mu \in \Irr (  U) }     \frac {p_\mu}{d_\mu} ~    I_{\mu}  \otimes \rho_\mu$,
with $\rho_\mu \in\Stset(\Cmplx^{m_\mu})$.  The square-root  purification of $\rho$ is then given by
\begin{equation*}
|\Psi\>  = \bigoplus_{\mu \in \Irr  (U) }   \sqrt{\frac{p_\mu}{d_\mu}}   |  I_{\mu}\>\!\>  |\rho_\mu^{\frac 12}\>\!\>  \in \bigoplus_{\mu \in \Irr  (U)}    \spc H_{\mu} \otimes \spc H_\mu  \otimes \Cmplx^{m_\mu}  \otimes \Cmplx^{m_\mu}.
\end{equation*}
Here the first copy of $\spc H_\mu$ is the representation space, while the remaining spaces contribute to the multiplicity. The action of the representation $\{U_g \otimes I_{\spc H}\}_{g \in \grp G}$ then generates the orbit $|\Psi_g\> : = \bigoplus_{\mu \in \Irr  (U) }   \sqrt{\frac{p_\mu}{d_\mu}} ~ | U^{\mu}_g\>\!\>  |\rho_\mu^{\frac 12}\>\!\> $.  This orbit can be obtained from  the orbit  $|\Phi_g\>  : =\bigoplus_{\mu \in \Irr  (U) }   \sqrt{\frac{p_\mu}{d_\mu}} ~ | U^{\mu}_g\>\!\> \in \widetilde{\spc H}$ by applying the isometry $V  :     \bigoplus_{\mu \in \Irr  (U_g)}    \spc H_{\mu} \otimes \spc H_\mu  \to \bigoplus_{\mu \in \Irr  (U)}    \spc H_{\mu} \otimes \spc H_\mu  \otimes \Cmplx^{m_\mu}  \otimes \Cmplx^{m_\mu}  $   defined by $V (  \bigoplus_{\mu \in \Irr  (U)}    |v_\mu\> )   =    \bigoplus_{\mu \in \Irr  (U)}   | v_\mu \>   |\rho_\mu^{\frac 12}\>\!\>$.  This proves that the state $|\Phi\>= \bigoplus_{\mu \in \Irr(U)}  \sqrt{\frac {p_\mu}{d_\mu}}  |I_{\mu} \>\!\>$ is an optimal input state. \qed

\subsection{Invariant cost functions and isotropic states: optimality of isotropic seeds}

Suppose that the cost function $c(\hat g, g)$ is \emph{invariant}, that is, both left- and right-invariant. Equivalently, this means that $c(\hat g, g)  = f (\hat g^{-1}  g)$ where $f$ is a class function.  From the discussion of the previous paragraphs we already know that  \emph{i)} the optimal input state  will be  class state and \emph{ii)} the optimal POVM will be covariant.  In fact, in this case the optimal POVM has an additional symmetry that reflects the symmetries of class states (see Eq. (\ref{symmetryclassstates})).   Furthermore, this additional property appears not only for class states but also for a more general family of input states that we call \emph{isotropic}.
\begin{defi}[Isotropic operators]
An operator $A \in \Lin (\widetilde{\spc H})$  with $\widetilde{\spc H}=\bigoplus_{\mu \in \Irr (U)}   \spc H_\mu \otimes \spc H_\mu$ is  \emph{isotropic} if
\begin{equation*}
\map U_g  \map U^R_g  ( A)  =  A  \qquad \forall g\in\grp G.
\end{equation*}
\end{defi}
The set of isotropic states (i.e. non-negative isotropic operators with unit trace) is convex.  Clearly, every class state is isotropic. As a consequence, every mixture of class states is an isotropic state.


\begin{theo}[Optimality of isotropic seeds]\label{theo:isotropicpovm}
Let $\grp G$ be a compact group, $\widetilde U$ be a  projective unitary representation of $\grp G$ on the Hilbert space  $\widetilde{\spc H}=\bigoplus_{\mu \in \Irr (U)}   \spc H_\mu \otimes \spc H_\mu$,  $c(\hat g,g)$ be an invariant cost function and $\rho  \in \Stset( \spc H)$ be an isotropic state. Then, the seed $\xi \in \Lin_+ (\widetilde{\spc H})$ of the optimal covariant measurement for the input state $\rho$  is isotropic.
\end{theo}

\Proof Since the cost function is left-invariant, the optimal POVM can be assumed to be covariant (theorem \ref{theo:optcovpovm}) and there is no difference between the average and worst case approaches. Moreover, for any covariant POVM $P(\d g)  =  \widetilde {\map U}_g (\xi)  \d g$ we can define another covariant POVM $P' (\d g)  = \widetilde{\map U}_g (\xi')  \d g$ with $\xi$ where $\xi'$ is isotropic.  Indeed, it is enough to choose $\xi' :=  \int_{\grp G}  \d h~  \widetilde{\map U}_h \widetilde{ \map U}_h^R  (\xi)$.  Note that  $P'(\d g)$  is still a resolution of the identity.  The uniform average cost of the strategy $(\rho, P')$ is the same of the strategy $(\rho, P)$:
\begin{align*}
c_{ave}  (\rho, P')  &=\int_{\grp G}  \d g \int_{\grp G}  \d \hat g  ~ c(\hat g, g)  \Tr[ \widetilde{\map U}_{\hat g} (\xi') \widetilde{ \map U}_g (\rho) ]\\
     &=\int_{\grp G}  \d g \int_{\grp G}  \d \hat g  \int_{\grp G}  \d h~ c(\hat g, g)  \Tr[ \widetilde{\map U}_{\hat g h} (\xi) \widetilde{ \map U}_g \widetilde{\map U}_h^{R \dag} (\rho) ]\\
 &=\int_{\grp G}  \d h \int_{\grp G}  \d g \int_{\grp G}  \d \hat g  ~ c(\hat g h, g h)  \Tr[ \widetilde{\map U}_{\hat g h} (\xi)  \widetilde{\map U}_{gh} (\rho) ]\\
 & = c_{ave}  (\rho,P),
  \end{align*}
having  used the fact that $\rho$ is isotropic (and hence $\widetilde{\map U}_h^{R\dag} (\rho)  =  \widetilde{\map U}_h (\rho)$) and the right-invariance of the cost function.
\qed

An example of isotropic seed is given by the \emph{class seed} $\xi  =  |\eta\>\<\eta|$, where $|\eta\> \in\spc H$ is the class vector
\begin{equation}\label{classseed}
|\eta\>  =  \bigoplus_{\mu \in\Irr (U)}  \sqrt{d_\mu}  |I\mu\>\!\>.
\end{equation}
We will now see that this seed plays a special role: any isotropic seed can be obtained from it by applying a suitable physical transformation.
\begin{theo}
$\xi \in\Lin_+ (\widetilde{\spc H})$ is an isotropic seed if and only if $\xi  =  \map C (|\eta\>\<\eta|)$  for some  completely positive map $\map C: \Lin (\widetilde{\spc H}) \to \Lin (\widetilde{\spc H})$ that is bistochastic (i.e.  trace-preserving and unit-preserving) and leaves all elements in the algebra $U'$ invariant.
\end{theo}
\Proof  Let us write $\xi  = \sum_{i=1}^r  |\eta_i\>\<\eta_i|$  and $|\eta_i\>  =  \bigoplus_{\mu \in \Irr (U) }  \sqrt{d_\mu}   |\eta_{i\mu} \>\!\>$. Defining  the operators
$C_i  : = \bigoplus_{\mu \in \Irr(U)}  \eta_{i\mu} \otimes I_\mu \in U'' $
and defining the completely positive map $\map C $ as $\map C(\rho) := \sum_{i=1}^r C_i \rho C_i^\dag$ we then obtain $\xi  = \map C (|\eta\>\<\eta|)$.
Clearly, one has $\map C (A) = A$ for every $A \in U'$.   The requirement that $\map C$ be bistochastic is equivalent to the requirement that $\xi$ be an isotropic seed.
First, we can prove that $\map C$ is unital:
\begin{align*}
\map C (I)  & =  \int_{\grp G} \d g ~   \map C  \widetilde{\map U}_g (|\eta\>\<\eta|)  =  \int_{\grp G} \d g  ~  \map C \widetilde{ \map U}^{R\dag}_g (|\eta\>\<\eta|) \\
& =  \int_{\grp G} \d g   ~ \widetilde{\map U}^{R \dag}_g  \map C  (|\eta\>\<\eta|)  =  \int_{\grp G} \d g ~ \widetilde {\map U}_g^{R \dag}  (\xi)  =  \int_{\grp G} \d g   ~ \widetilde{\map U}_g (\xi) = I
\end{align*}
To prove that $\map C$ is trace-preserving, we equivalently prove that $\map C^\dag$ is unital.  Using Eq. (\ref{sguscia}) it is immediate to obtain $\map C^{R\dag} (I) = I$:  indeed, we have
\begin{align*}
\map C^{R\dag} (I)  & =  \int_{\grp G} \d g ~   \map C^{R\dag}  \widetilde{\map U}_g (|\eta\>\<\eta|)  =  \int_{\grp G} \d g  ~  \widetilde{\map U}_g  \map C^{R\dag} (|\eta\>\<\eta|) \\
& =  \int_{\grp G} \d g   ~ \widetilde{\map U}_g  \map C  (|\eta\>\<\eta|)  =  \int_{\grp G} \d g   ~\widetilde{ \map U}_g (\xi)  = I
\end{align*}
Finally, it is enough to observe that $\map C^\dag  =  \map S  \left(\map C^{R \dag }\right)^* \map S$ (see Eq. (\ref{swap})) and, therefore,  $\map C^\dag (I)  =  I$.\qed

\begin{cor}\label{muoio}
Let $\xi \in\Lin (\spc H)$ be an isotropic seed, written as $\xi =  \sum_{i=1}^r  |\eta_i\>\<\eta_i|$, $|\eta_i \>  = \bigoplus_{\mu \in \Irr (U)}  \sqrt{d_\mu}  |\eta_{i\mu}\>\!\>$.
Then we have  $\sum_{i=1}^r  \eta_{i\mu}  \eta_{i\mu}^\dag  =  \sum_{i=1}^r  \eta_{i\mu}^T  \eta_{i\mu}^* = I_\mu$.
\end{cor}

In the following the covariant POVM $E(\d \hat g)  =  \widetilde{ \map U}_{\hat g}  (|\eta\>\<\eta|) \d \hat g$ generated by the class vector $|\eta\>$ in Eq. (\ref{classseed})  will be called \emph{class POVM}.

\begin{cor}
Any covariant POVM $P(\d \hat g) = \widetilde{\map U}_g (\xi)$ with isotropic seed can be obtained from the class POVM $ E (\d \hat g)  =  \widetilde{\map U}_{\hat g} (|\eta\>\<\eta|) \d \hat g$ by application of a bistochastic map $\map C^{R\dag}$.
\end{cor}
Note that the bistochastic map $\map C^{R\dag}$ is \emph{covariant}, that is $\widetilde{\map U}_g \map C^{R\dag}  = \map C^{R\dag}  \widetilde{\map U}_g \forall g \in \grp G$.

\begin{cor}
Let $\rho \in\Stset(\widetilde{\spc H})$ be an isotropic state and $P(\d \hat g) = \widetilde{\map U}_g (\xi)$ be a covariant POVM with isotropic seed. Then, the estimation strategy $(\rho, P)$ has the same cost of the estimation strategy $(\map C^{R\dag} (\rho), E)$, where $E$ is the class POVM $E(\d\hat g)  = \widetilde{ \map U}_{\hat g} (|\eta\>\<\eta|) \d \hat g$.
\end{cor}

The question now is: why should we perform a physical transformation on the input state $\rho$  before performing the measurement?  Intuitively, one might expect that the application of the transformation $\map C^{R\dag}$ can only degrade the information contained in the signal states $\rho_g := \widetilde{\map U}_g (\rho)$. In the next paragraph I will show that the intuition is correct in the case where $\rho$ is a class state and the cost function $c (\hat g, g)$ has negative Fourier transform.

\subsection{Invariant cost functions with negative Fourier transform: optimality of  square-root measurements}

Here we will confine our attention to a special family of invariant cost functions, that is, to cost functions of the form
\begin{equation}\label{nicecost}
c(\hat g, g)  =  \sum_{\sigma \in \UIrr (\grp G)} a_\sigma  \chi^*_\sigma (\hat g^{-1} g)    \qquad  a_\sigma  \le 0  \quad \forall  \sigma \in \UIrr (\grp G) .
\end{equation}
where $\UIrr (\grp G)$ denotes the set of all unitary irreducible representations of $\grp G$
(the complex conjugate of the character has been introduced   just for later convenience).
Optimization problems with cost functions in this family are  the maximization of the \emph{entanglement fidelity}, corresponding to the cost function given by $c(\hat g, g) = -|\Tr[V_{\hat g^{-1} g}] /n|^2$, where $V: \grp G  \to \Lin (\Cmplx^n)$ is a unitary representation of $\grp G$, and the \emph{maximum likelihood} \cite{ml}, corresponding to the cost function $c(\hat g, g) = - \delta(\hat g^{-1} g)$.

 Incidentally, we observe that each cost function of the form in Eq. (\ref{nicecost}) defines a positive semidefinite product   on the vector space of class functions,  given by
\begin{align*}
\<  f_1, f_2\>_c  &: = - \int_{\grp G} \d \hat g  \int_{\grp G} \d g ~  c(\hat g, g)   f_1^* (\hat g)  f_2(g)\\
 & =  -  \sum_{\sigma \in \UIrr (\grp G)}   a_\sigma    f_{1,\sigma}^*  f_{2,\sigma}.
\end{align*}

The next theorem summarizes  the structures presented so far:

\begin{theo}\label{theo:optsquareroot} {\bf Optimal estimation strategy for invariant functions with negative Fourier transform.}
Let $\grp G$ be a compact group, $\widetilde U$ be a unitary projective representation of $\grp G$ on the Hilbert space $\widetilde{\spc H} =\bigoplus_{\mu \in \Irr (U)}   \spc H_\mu \otimes \spc H_\mu$ and $c(\hat g,g)$ be an invariant cost function of the form of Eq. (\ref{nicecost}). Then, the class POVM $E (\d \hat g) = \map U_g (|\eta\>\<\eta|) \d \hat g  $ is optimal for every class state $|\Phi\>$ of the form $|\Phi\>  = \bigoplus_{\mu \in \Irr (U) }  \sqrt{p_\mu/d_\mu}  |I_\mu\>\!\>$.
The cost is given by
$c_{ave}   =  \sum_{\sigma \in \UIrr (\grp G)}   a_\sigma m_{\sigma}^{\mu\nu^*}   \sqrt{p_\mu p_\nu},$
where  $m_{\sigma}^{\mu\nu^*}$ is the multiplicity of the irreducible representation $U^\sigma$ in  the tensor product $U^\mu \otimes  U^{\nu*}$.

\end{theo}
\Proof  By theorem \ref{theo:isotropicpovm}  we can assume without loss of generality that the optimal POVM is covariant, i.e. $P(\d \hat g)  = \map U_g (\xi) \d \hat g$, and that the seed $\xi$ is isotropic.  Writing the seed as $\xi = \sum_{i=1}^r  |\eta\>\<\eta|$ with $|\eta\>  = \bigoplus_{\mu \in \Irr (U)}  \sqrt{d_\mu}    |\eta_{i \mu\>\!\>}$  we obtain for the cost
\begin{align}
\nonumber c ( P, \Phi| e)  & = \int_{\grp G} \d \hat g ~  c(\hat g , e)  \<  \Phi|   \widetilde{\map U}_{\hat g}  (\xi)  |\Phi\>  \\
\nonumber &= \sum_{\sigma \in \UIrr (\grp G)}   \sum_{\mu,\nu  \in  \Irr (U)}  a_\sigma \sqrt{\frac{p_\mu p_\nu}{d_\mu d_\nu}}    \int_{\grp G} \d \hat g ~   \chi^*_\sigma (\hat g) ~ \<\!\<  I_\mu|   \widetilde{\map U}_{\hat g}  (\xi)  |I_\nu\>\!\>  \\
 \nonumber &= \sum_{\sigma \in \UIrr (\grp G)}   \sum_{\mu,\nu  \in  \Irr (U)}  a_\sigma \sqrt{p_\mu p_\nu}  \sum_{i=1}^r   \int_{\grp G} \d \hat g ~   \chi_\sigma^* (\hat g) ~ \<\!\<  I_\mu|   (U_{\hat g}^\mu  \otimes I_\mu)  |\eta_{i\mu}\>\!\>     \<\!\<  \eta_{i\nu} |  (U^{\nu \dag}_{\hat g}  \otimes I_\nu)   |I_\nu\>\!\>  \\
 \nonumber &= \sum_{\sigma \in \UIrr (\grp G)}   \sum_{\mu,\nu  \in  \Irr (U)}  a_\sigma  c\sqrt{p_\mu p_\nu}    \sum_{i=1}^r   \int_{\grp G} \d \hat g ~   \chi^*_\sigma (\hat g) ~ \<\!\<  I_\mu | U_{\hat g}^{\mu}   \eta_{i\mu}\>\!\>     \<\!\<   U^\nu_{\hat g} \eta_{i\nu} |  I_\nu\>\!\>  \\
 \nonumber &= \sum_{\sigma \in \UIrr (\grp G)}   \sum_{\mu,\nu  \in  \Irr (U)}  a_\sigma  \sqrt{p_\mu p_\nu}   \sum_{i=1}^r   \int_{\grp G} \d \hat g ~   \chi^*_\sigma (\hat g) ~ \Tr[ U_{\hat g}^{\mu}   \eta_{i\mu}  \otimes    U^{\nu*}_{\hat g} \eta^*_{i\nu} ]  \\
 &= \sum_{\sigma \in \UIrr (\grp G)}   \sum_{\mu,\nu  \in  \Irr (U)}  \frac{a_\sigma  \sqrt{p_\mu p_\nu} }{d_\sigma}  \sum_{i=1}^r    \Tr[ \Pi_{\sigma}^{\mu\nu^*}   (\eta_{i\mu}  \otimes   \eta^*_{i\nu}) ] , \label{prima}
 \end{align}
where $P_{\sigma}^{\mu \nu^*}$  is the projector on the direct sum of all irreducible subspaces of $\spc H_\mu \otimes \spc H_\nu$ that carry the irreducible representation $U^\sigma$ as a subrepresentation of the tensor product  $U^{\mu}\otimes  U^{\nu*}$.
Since $a_\sigma \le 0$ for every $\sigma$, we have the bound
\begin{align*}
c ( P, \Phi|e) &\le \sum_{\sigma \in \UIrr (\grp G)}   \sum_{\mu,\nu  \in  \Irr (U)}  \frac{a_\sigma \sqrt{p_\mu p_\nu} } {d_\sigma}     \left|\sum_{i=1}^r    \Tr[ \Pi_{\sigma}^{\mu\nu^*}   (\eta_{i\mu}  \otimes   \eta^*_{i\nu}) ] \right|,\\
& =  \sum_{\sigma \in \UIrr (\grp G)}   \sum_{\mu,\nu  \in  \Irr (U)}  \frac{a_\sigma\sqrt{p_\mu p_\nu} } {d_\sigma}      \left|\sum_{i=1}^r    \Tr[ \Pi_{\sigma}^{\mu\nu^*}   (\eta_{i\mu}  \otimes   I_\nu )( I_\mu \otimes  \eta^*_{i\nu})  \Pi_{\sigma}^{\mu\nu^*}  ] \right|,\\
& \le \sum_{\sigma \in \UIrr (\grp G)}   \sum_{\mu,\nu  \in  \Irr (U)}  \frac{a_\sigma\sqrt{p_\mu p_\nu} } {d_\sigma}     \sqrt{ \sum_{i=1}^r    \Tr[ \Pi_{\sigma}^{\mu\nu^*}  (\eta_{i\mu}  \eta_{i\mu}^\dag \otimes I_\nu) ]}  \sqrt{ \sum_{j =1}^r\Tr [       (I_\nu \otimes   \eta^{T}_{i\nu}  \eta^*_{i\nu})  \Pi_{\sigma}^{\mu\nu^*}  ] },\\
\end{align*}
having used the Cauchy-Schwarz inequality  $ |\Tr  [A^\dag B]  | \le \sqrt{\Tr[A^\dag A]  \Tr[B^\dag B]  }$ for block matrices $A = \bigoplus_{i=1}^r  A_i$ and $B = \bigoplus_{i=1}^r  B_i$.    Finally, using the normalization conditions of corollary \ref{muoio}  we obtain
\begin{align*}
c ( P, \Phi,e)  \le \sum_{\sigma \in \UIrr (\grp G)}   \sum_{\mu,\nu  \in  \Irr (U)}  \frac{a_\sigma \sqrt{p_\mu p_\nu} } {d_\sigma}        \Tr [\Pi_{\sigma}^{\mu\nu^*}  ]  \le \sum_{\sigma \in \UIrr (\grp G)}   \sum_{\mu,\nu  \in  \Irr (U)}  a_\sigma    m_{\sigma}^{\mu\nu^*}  \sqrt{p_\mu p_\nu}      ,
\end{align*}
where $m_{\sigma}^{\mu\nu^*}$ is the multiplicity of $U^\sigma$ in the tensor product of $U^\mu\otimes U^{\nu*}$.
On the other hand, by substituting in Eq. (\ref{prima}) it is immediate to see that the class POVM $E (\d \hat g)  = \map U_{\hat g}  (|\eta\>\<\eta|)$ achieves the bound and hence is the best possible POVM. \qed

The optimality of the class POVM can be rephrased as a result about the optimality of \emph{square-root measurements}  \cite{squareroot} (see also \cite{belavkin} for a surprising \emph{ante litteram} example). The general definition of square-root measurement that we will consider here is the following:
\begin{defi}
Let $(X,\sigma(X))$ be a measurable space and let  $\rho(\d x) : \sigma (X) \to \Lin_+ (\spc H),  B  \mapsto \rho_B$ be an ensemble.  The square-root measurement associated to $\rho$ is the POVM $P:  \sigma (B)  \mapsto \Lin_+ (\Supp (\rho_X))$ given by
$P_B  : = \rho_X^{- \frac 12}  \rho_B  \rho_X^{-\frac 12}$, where $\rho_X^{-\frac 12}$ is the inverse of $\rho_X^{\frac 12}$ on its support.
\end{defi}
It is easy to see that the class POVM  $E(\d \hat g) = \widetilde{\map U}_{\hat g} (|\eta\>\<\eta|)  \d \hat g$ is the square-root measurement associated to the ensemble $\rho(\d  g)  : = |\Phi_g\>\<\Phi_g| \d g$ where $|\Phi_g\> = \widetilde U_g|\Phi\>$ for some class state $|\Phi\>= \bigoplus_{\mu \in \Irr (U)}  \sqrt{\frac{p_\mu}{d_\mu}}  |I_{\mu}\>\!\>$. Indeed, we have
$\rho_\grp G  = \int_{\grp G}  \d g ~  |\Phi_g\>\<\Phi_g|  = \bigoplus_{\mu \in \Irr (U)}  \frac{p_\mu}{d_\mu^2} ~ I_\mu \otimes I_\mu,$
and therefore
$\rho_\grp G^{-\frac 12 } |\Phi_{\hat g}\>  = \widetilde U_{\hat g}  \rho_{\grp G}^{-\frac 12}   |\Phi\>  = \widetilde U_{\hat g}  \left(   \sum_{\mu \in \Irr (U)}  \sqrt{{d_\mu}}  |I_\mu\>\!\>  \right)  = \widetilde U_{\hat g} |\eta\>,$
which means  $ E(\d \hat g)  =   \rho_{\grp G}^{- \frac 12} |\Phi_{\hat g}\>\<\Phi_{\hat g}|  \rho_{\grp G}^{- \frac 12}  \d \hat g$.

A slight generalization of Theorem \ref{theo:optsquareroot} is then given by the following:
\begin{cor}\label{cor:optsquareroot}
Let $\grp G$ be a compact group, $\widetilde U$ be a  projective unitary representation of $\grp G$ on the Hilbert space $\widetilde{\spc H}  =\bigoplus_{\mu \in \Irr (U)}   \spc H_\mu \otimes \spc H_\mu$ and $c(\hat g,g)$ be an invariant cost function of the form of Eq. (\ref{nicecost}). Then, the square-root measurement is optimal for every state $|\Psi\> \in \widetilde{\spc H}$ of the form $ |\Psi\> = V  |\Phi\>$ where $|\Phi\>\in\widetilde {\spc H}$ is a class state as in Eq. (\ref{optinp}) and $V\in\Lin(\widetilde {\spc H})$ is a unitary matrix in $U'$.
\end{cor}
\Proof The square-root measurement associated to the ensemble $|\Psi_g\>\<\Psi_g| \d g$ is $P(\d \hat g)  =  \map U_g  \map V (|\eta\>\<\eta|)$.  It is easy to see that the cost of the strategy $(\Psi, P)$ equal the cost of  the strategy $(\Phi, E)$: indeed, the conditional probabilities are the same:
\begin{align*}
 p'(\d \hat g|  g)&: = \Tr[  P(\d \hat g)  \map U_g (|\Psi\>\<\Psi|)  \\
 &= \Tr[ \map U_{\hat g}  \map V  (|\eta\>\<\eta|)  \map U_g  \map V(|\Phi\>\<\Phi)] \d \hat g\\
  &= \Tr[\map U_{\hat g}  (|\eta\>\<\eta| )  \map U (|\Phi\>\<\Phi|)] \d \hat g := p(\d \hat g|g)
 \end{align*}
 This means that the square-root measurement $P(\d\hat g)$ must be optimal for $|\Psi\>$, otherwise $E$ would not be optimal for $|\Phi\>$.
 \qed

\section{Conclusions and open problems}
This paper provided a synthesis of the main group theoretical structures appearing in the estimation of unitary transformations on finite dimensional Hilbert spaces. I deliberately avoided any mention to infinite dimensions, in order to keep the exposition at the simplest possible level and to highlight the structures that underly optimal estimation in the neatest possible way.  The development of a general theory of optimal estimation for infinite dimensional systems is mostly an open problem, especially in the case of non-compact groups, for which there are many particular results (especially for abelian groups) but still no unifying framework. The development of a general theory of estimation for non-abelian non-compact groups rests as a stimulating challenge for future research.

\section{Acknowledgments}  This paper contains an advanced elaboration of some ideas present in my PhD thesis ``Optimal estimation of quantum signals in the presence of symmetry" (Pavia, October 2006).  I would like to take this opportunity to express my gratitude to my PhD advisor and friend G. M. D'Ariano and to my extraordinary collaborators P. Perinotti and M. F. Sacchi.
Research at Perimeter Institute for Theoretical Physics is supported in part by the Government of Canada through NSERC and by the Province of Ontario through MRI.

\end{document}

%% file: myQcircuit.tex
\usepackage[matrix,frame,arrow]{xy}
\usepackage{amsmath}

\newcommand{\qw}[1][-1]{\ar @{-} [0,#1]}

\newcommand{\gate}[1]{*{\xy *+<.6em>{#1};p\save+LU;+RU **\dir{-}\restore\save+RU;+RD **\dir{-}\restore\save+RD;+LD **\dir{-}\restore\POS+LD;+LU **\dir{-}\endxy} \qw}
\newcommand{\Qcircuit}[1][0em]{\xymatrix @*[o] @*=<#1>}

\newcommand{\pureghost}[1]{*+<1em,.9em>{\hphantom{#1}}}
\newcommand{\multiprepareC}[2]{*+<1em,.9em>{\hphantom{#2}}\save[0,0].[#1,0];p\save !C
  *{#2},p+RU+<0em,0em>;+LU+<+.8em,0em> **\dir{-}\restore\save +RD;+RU **\dir{-}\restore\save
  +RD;+LD+<.8em,0em> **\dir{-} \restore\save +LD+<0em,.8em>;+LU-<0em,.8em> **\dir{-} \restore \POS
  !UL*!UL{\cir<.9em>{u_r}};!DL*!DL{\cir<.9em>{l_u}}\restore}
\newcommand{\prepareC}[1]{*{\xy*+=+<.5em>{\vphantom{#1\rule{0em}{.1em}}}*\cir{l^r};p\save*!L{#1} \restore\save+UC;+UC+<.5em,0em>*!L{\hphantom{#1}}+R **\dir{-} \restore\save+DC;+DC+<.5em,0em>*!L{\hphantom{#1}}+R **\dir{-} \restore\POS+UC+<.5em,0em>*!L{\hphantom{#1}}+R;+DC+<.5em,0em>*!L{\hphantom{#1}}+R **\dir{-} \endxy}}
\newcommand{\poloFantasmaCn}[1]{{{}^{#1}_{\phantom{#1}}}}

%% file: IGTMPchiribella.bbl
\medskip
\begin{thebibliography}{99}
\bibitem{metro}  D. Leibfried \emph{et al},  Nature {\bf 438}, 639 (2005). T. Nagata, R. Okamoto, J. L. OÕBrien, K. Sasaki, and S. Takeuchi, Science {\bf 316}, 726 (2007).  B. L. Higgins, D. W. Berry, S. D. Bartlett, H. M. Wiseman, G. J. Pryde,   Nature {\bf 450}, 393 (2007).
\bibitem{aag}   S. T. Ali, J. P. Antoine and J. P. Gazeau, \emph{Coherent States, Wavelets and Their Generalizations}, Springer-Verlag, 2000.
\bibitem{clocks} V. Bu\u zek, R. Derka, and S. Massar, Phys. Rev. Lett. {\bf 82}, 2207, (1999).
\bibitem{refframe} G. Chiribella, G. M. D'Ariano, P. Perinotti, and M. F. Sacchi, Phys. Rev. Lett. {\bf 93}, 180503 (2004).
\bibitem{review}   S. D. Bartlett, T. Rudolph, R. W. Spekkens Rev. Mod. Phys. {\bf 79}, 555 (2007).
\bibitem{dense} C. H. Bennett and S. J. Wiesner, Phys. Rev. Lett. {\bf 69}, 2881 (1992).
\bibitem{sigmaalg}  Precisely, $\sigma (\grp G)$ denotes the Borel $\sigma$-algebra generated by the topology of $\grp G$.
\bibitem{PLA} G. M. D'Ariano, P. Lo Presti, and M. F. Sacchi,
Phys. Lett. A {\bf 272}, 32 (2000).
\bibitem{folland} G. B. Folland, editor. \emph{A course in abstract harmonic analysis}. Studies in Advanced Mathematics. CRC Press, Boca Raton, FL, 1995.
\bibitem{dirac} The definition is more intuitive if we use formally the Dirac vectors $\{|h\>\}_{h \in \grp G}$, with $\< h_1 |h_2\> = \delta(h_1^{-1} h_2)$ and write $|f\>  =\int_{\grp G} \d h ~  f(h)  |h\>$.  With this notation we have $T_g|h\>  = \omega (g,h)  |gh\>$, which shows clearly that $T$ is a projective representation with multiplier $\omega$.
\bibitem{helstrom} C. W. Helstrom, {\it Quantum detection and
    estimation theory} (Academic Press, New York, 1976).
\bibitem{holevo} A. S. Holevo, \emph{Probabilistic and Statistical
   Aspects of Quantum Theory} (North Holland, Amsterdam, 1982).
\bibitem{belavkin} V. P. Belavkin and V. P. Maslov, \emph{Design of Optimal Dynamic Analyzers: Mathematical Aspects of Wave Pattern Recognition}, Mathematical Aspects Of Computer Engineering Advances in Science and Technology in the USSR, Mir Publisher (1988), http://arxiv.org/abs/quant-ph/041203.
\bibitem{entest}  G. Chiribella, G. M. D'Ariano, and M. F. Sacchi, Phys. Rev. A {\bf 72}, 042338 (2005).
\bibitem{purification}  G. Chiribella, G. M. D'Ariano, and P. Perinotti,  Phys. Rev. A {\bf 81}, 062348 (2010).
\bibitem{schro} E. Schr\"odinger, Proc. Camb. Phil. Soc. {\bf 31}, 555
 (1935).
\bibitem{ml} G. Chiribella, G. M. D'Ariano, P. Perinotti, M. F. Sacchi,  Int. J. Quantum Inf. {\bf 4}, 453 (2006).
\bibitem{squareroot} P. Hausladen and W. K. Wootters, J. Mod. Opt.
  {\bf 41}, 2385 (1994).
\end{thebibliography}
